\documentclass{Interspeech2024}
\usepackage{multirow}
\usepackage{booktabs}
\usepackage{glossaries}
\usepackage{bbold}
\usepackage{subdepth}  
\usepackage{subcaption}
\usepackage[moderate]{savetrees}
\usepackage{nicefrac}
\usepackage{cite}
\usepackage{microtype}  

\newcommand*{\bPhi}[1][]{\ensuremath{\boldsymbol{\Phi}^{#1}_{f,t}}}

\newcommand*{\tran}{\ensuremath{{}^{\mkern-1.5mu\mathsf{T}}}}

\newcommand*{\hermconj}{\ensuremath{{}^{\mathsf{H}}}}

\newacronym{stft}{STFT}{short-time Fourier transform}
\newacronym{mvdr}{MVDR}{minimum variance distortionless response}
\newacronym{dnn}{DNN}{deep neural network}
\newacronym{pesq}{PESQ}{perceptual evaluation of speech quality}
\newacronym{estoi}{ESTOI}{extended short-term objective intelligibility}
\newacronym{snr}{SNR}{signal-to-noise-ratio}
\newacronym{sisdr}{SI-SDR}{scale-invariant signal-to-distortion-ratio}
\newacronym{sdr}{SDR}{signal-to-distortion-ratio}
\newacronym[longplural={spatial covariance matrices}]{scm}{SCM}{spatial covariance matrix}
\newacronym{iscm}{ISCM}{instantaneous SCM}
\newacronym{tac}{TAC}{transform-average-concatenate}
\newacronym{ipd}{IPD}{inter-microphone phase difference}
\newacronym{ild}{ILD}{inter-microphone level difference}
\newacronym{mha}{MHA}{multi-head attention}
\newacronym{sha}{SHA}{single-head attention}
\newacronym{rir}{RIR}{room impulse response}
\newacronym{asa}{ASA}{attention-based SCM aggregator}
\newacronym{asr}{ASR}{automatic speech recognition}
\newacronym{wsj0}{WSJ0}{Wall Street Journal}

\title{Array Geometry-Robust Attention-Based Neural Beamformer for Moving Speakers}

\name[affiliation={2*}]{Marvin}{Tammen}
\name[affiliation={1}]{Tsubasa}{Ochiai}
\name[affiliation={1}]{Marc}{Delcroix}
\name[affiliation={1}]{Tomohiro}{Nakatani}
\name[affiliation={1}]{Shoko}{Araki}
\name[affiliation={2}]{Simon}{Doclo}

\address{
  $^1$NTT Corporation, Japan;
  $^2$Carl von Ossietzky Universität Oldenburg, Germany
  }
\email{marvin.tammen@uol.de}


\interspeechcameraready

\begin{document}
\maketitle

\renewcommand{\thefootnote}{\fnsymbol{footnote}}
\footnotetext[1]{This work was done during an internship at NTT Corporation.}
\renewcommand{\thefootnote}{\arabic{footnote}}  
\setcounter{footnote}{1}

\begin{abstract}
Although mask-based beamforming is a powerful speech enhancement approach, it often requires manual parameter tuning to handle moving speakers.
Recently, this approach was augmented with an attention-based spatial covariance matrix aggregator (ASA) module, enabling accurate tracking of moving speakers without manual tuning.
However, the deep neural network model used in this module is limited to specific microphone arrays, necessitating a different model for varying channel permutations, numbers, or geometries.
To improve the robustness of the ASA module against such variations, in this paper we investigate three approaches: training with random channel configurations, employing the transform-average-concatenate method to process multi-channel input features, and utilizing robust input features.
Our experiments on the CHiME-3 and DEMAND datasets show that these approaches enable the ASA-augmented beamformer to track moving speakers across different microphone arrays unseen in training.
\end{abstract}


\section{Introduction}
In many speech communication applications, the microphone signals are corrupted by ambient noise, reducing speech quality and intelligibility as well as degrading the performance of \gls{asr} systems.
When multiple microphones are available, good noise reduction performance with low speech distortion can be achieved using beamforming, provided that accurate estimates of the required \glspl{scm} are available~\cite{docloMultichannelSignalEnhancement2015,gannotConsolidatedPerspectiveMultimicrophone2017}. 

In mask-based beamformers, the \gls{scm} estimation task has often been offloaded to \glspl{dnn}~\cite{heymannNeuralNetworkBased2016,higuchiRobustMVDRBeamforming2016,erdoganImprovedMVDRBeamforming2016,boeddekerExploringPracticalAspects2018,wangMultimicrophoneComplexSpectral2021,casebeerNICEBeamNeuralIntegrated2021a,zhangADLMVDRAllDeep2021,wangAttentionDrivenMultichannelSpeech2023a,jukicFlexibleMultichannelSpeech2023,tammenParameterEstimationProcedures2023,ochiaiMaskBasedNeuralBeamforming2023}.
These beamformers have demonstrated remarkable performance in recent \gls{asr} challenges such as the CHiME-4 challenge, while typically being applicable to arbitrary channel configurations, i.e., to arbitrary permutations and numbers of channels as well as associated microphone array geometries.
However, most studies have focused on stationary acoustic scenarios, where the \glspl{scm} are estimated across entire utterances~\cite{heymannNeuralNetworkBased2016,erdoganImprovedMVDRBeamforming2016,wangMultimicrophoneComplexSpectral2021,jukicFlexibleMultichannelSpeech2023}.
This approach falls short in realistic acoustic scenarios involving moving speakers, where the \glspl{scm} are inherently time-varying.
Various heuristic tracking methods, such as block-online estimation or recursive smoothing, have been proposed~\cite{boeddekerExploringPracticalAspects2018}.
However, these methods heavily rely on manual tuning of parameters such as forgetting factors, which are highly dependent on the acoustic scenario, potentially leading to poor tracking performance.

To avoid such manual tuning and achieve a better tracking performance, a mask-based beamformer employing an \gls{asa} module has been proposed in \cite{ochiaiMaskBasedNeuralBeamforming2023}.
The \gls{asa} module temporally aggregates instantaneous estimates of the \glspl{scm} to compute time-varying speech and noise \glspl{scm}.
In \cite{ochiaiMaskBasedNeuralBeamforming2023}, it was demonstrated that the \gls{asa} module accurately tracks moving speakers, outperforming heuristic tracking methods.
However, since the employed training procedure, \gls{dnn} architecture, and input features depend on the channel configuration, the mask-based beamformer with \gls{asa} lost the ability to operate with arbitrary microphone array configurations, one of the key benefits of conventional mask-based beamformers.
\begin{figure}[t]
  \centering
  \includegraphics[width=228pt]{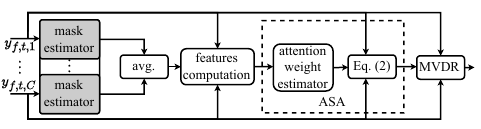}
  \caption{Overview of mask-based \acrshort{mvdr} beamformer with \gls{asa}. Grey vertically stacked boxes share weights.}
  \vspace{-12pt}
  \label{fig: overview}
\end{figure}

Aiming at realizing a mask-based beamformer with \gls{asa}
for arbitrary microphone arrays, in this paper we propose three approaches extending the prior work in \cite{ochiaiMaskBasedNeuralBeamforming2023}.
First, we investigate incorporating random channel configurations in the training procedure to prevent the \gls{dnn} from overfitting to specific channel permutations and channel numbers.
Second, we propose to employ the \gls{tac} method~\cite{luoEndtoendMicrophonePermutation2020} in the \gls{asa} module to process multi-channel features, allowing for any channel number and enabling permutation invariance.
The \gls{tac} method was originally proposed for channel permutation-invariant multi-channel source separation and has been successfully employed, e.g., in time-frequency masking algorithms~\cite{yoshiokaVarArrayArrayGeometryAgnosticContinuous2022,wangNeuralSpeechSeparation2020} and stationary mask-based beamformers~\cite{jukicFlexibleMultichannelSpeech2023}.
Third, we investigate utilizing input features that are less sensitive to variations of the channel configuration than the input features in \cite{ochiaiMaskBasedNeuralBeamforming2023}. 
Through experiments on the CHiME-3~\cite{barkerThirdCHiMESpeech2015} and DEMAND~\cite{thiemannDiverseEnvironmentsMultichannel2013} datasets including moving speakers, we demonstrate the benefit of jointly integrating the three proposed approaches into the \gls{asa} module.
Notably, our proposed approaches not only maintain high performance under matched conditions but also yield a good speech enhancement performance even for microphone arrays unseen during training, consistently outperforming a baseline mask-based beamformer with recursive smoothing and the mask-based beamformer with the original \gls{asa} in \cite{ochiaiMaskBasedNeuralBeamforming2023}.

\section{Mask-Based Beamformer With Attention-Based SCM Aggregator}
In this section, we provide an overview of the mask-based beamformer with \gls{asa}~\cite{ochiaiMaskBasedNeuralBeamforming2023}, depicted in Fig. \ref{fig: overview}.
The \gls{mvdr} beamformer is described in Section \ref{sec: mvdr}, the estimation of the required \glspl{scm} and the time-frequency masks is described in Section \ref{sec:scm}, and the computation of the features and the \gls{asa} module are described in Section \ref{sec: attention weight estimation}.

\subsection{MVDR Beamformer}
\label{sec: mvdr}
We consider an acoustic scenario with a single moving speaker and additive noise in a reverberant room, recorded by a set of $C$ microphones. 
In the \gls{stft} domain, the vector comprising the $C$ noisy microphone signals can be written as $\mathbf{y}_{f,t} = [ y_{f,t,c=1}, \ \dots, \ y_{f,t,c=C}]\tran \in \mathbb{C}^C$, where $f$, $t$, and $c$ denote the frequency bin index, the time frame index, and the channel index, respectively, and $\cdot\tran$ denotes the transpose operator.
Assuming that the (time-varying) acoustic transfer function between the speaker and the microphones is shorter than the \gls{stft} frame length, the noisy vector can be written as $\mathbf{y}_{f,t} = \mathbf{h}_{f,t} s_{f,t} + \mathbf{n}_{f,t}$, where $\mathbf{h}_{f,t} \in \mathbb{C}^C$, $s_{f,t} \in \mathbb{C}$, and $\mathbf{n}_{f,t} \in \mathbb{C}^C$ denote the acoustic transfer function, the speech source, and the additive noise component, respectively.

In beamforming approaches, the target speech component $x_{f,t,c=r} = h_{f,t,c=r} s_{f,t}$ at a reference microphone $r$ is typically estimated by applying a linear filter $\mathbf{w}_{f,t} \in \mathbb{C}^C$ to the noisy vector, i.e., $\widehat{x}_{f,t,r} = \mathbf{w}\hermconj_{f,t} \mathbf{y}_{f,t}$,
where $\cdot\hermconj$ denotes the conjugate transpose operator.
Aiming at minimizing the output noise power spectral density while leaving the target speech component undistorted, the \gls{mvdr} beamformer can be derived as~\cite{soudenOptimalFrequencyDomainMultichannel2010}:
\begin{equation}
    \label{eq: mvdr}
    \mathbf{w}_{f,t} = \frac{\left( \bPhi[n] \right)^{-1} \bPhi[x]}{\mathrm{tr} \left( \left( \bPhi[n] \right)^{-1} \bPhi[x] \right)} \mathbf{u}_r,
\end{equation}
where $\bPhi[x] \in \mathbb{C}^{C \times C}$ and $\bPhi[n] \in \mathbb{C}^{C \times C}$ denote the speech and noise \glspl{scm}, respectively, $\mathrm{tr} \left( \cdot \right)$ denotes the trace operator, and $\mathbf{u}_r \in \{0, 1\}^C$ denotes a selection vector with a 1 as the $r$-th element and 0 otherwise.

\subsection{Spatial Covariance Matrix Estimation}
\label{sec:scm}
To implement the \gls{mvdr} beamformer in \eqref{eq: mvdr}, estimates of the speech and noise \glspl{scm} $\bPhi[x]$ and $\bPhi[n]$ are required.
To allow estimating time-varying \glspl{scm}, in \cite{ochiaiMaskBasedNeuralBeamforming2023} the following temporal aggregation mechanism has been proposed:
\begin{align}
    \widehat{\boldsymbol{\Phi}}_{f,t}^\nu &= \sum_{\tau=1}^T a_{t, \tau}^\nu \underbrace{m_{f,\tau}^\nu \mathbf{y}_{f,\tau} \mathbf{y}\hermconj_{f,\tau}}_{= \widehat{\boldsymbol{\Psi}}^\nu_{f,\tau}}, \label{eq: iscm} 
\end{align}
where $\nu \in \{ x, n \}$ indicates the speech or noise component, $m^\nu_{f,t}$ denotes a time-frequency mask, $\widehat{\boldsymbol{\Psi}}_{f,\tau}^\nu \in \mathbb{C}^{C \times C}$ is an \gls{iscm} estimate, and $T$ denotes the number of time frames.
The frequency-independent attention weights $\mathbf{a}_t^\nu = \left[a_{t, \tau=1}^\nu, \ \ldots, \ a_{t, \tau=T}^\nu\right]\tran \in \mathbb{R}^T$ control how the \gls{iscm} estimates are temporally aggregated to yield estimates of the speech and noise \glspl{scm} at time frame $t$.
The time-frequency mask is typically obtained by applying a \gls{dnn}-based mask estimator independently for each channel, followed by averaging across channels, i.e., $m^\nu_{f,t} = \frac{1}{C} \sum^C_{c=1} m^\nu_{f,t,c}$.

\subsection{Attention Weight Estimation}
\label{sec: attention weight estimation}
\begin{figure}[t]
    \centering
    \includegraphics[width=\linewidth]{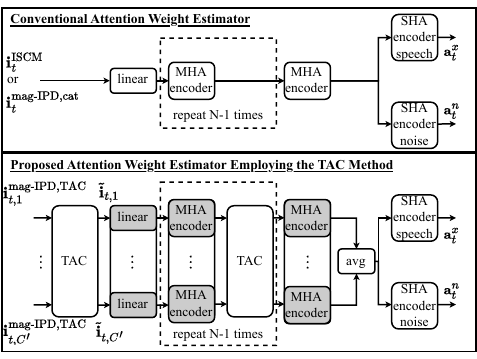}
    \caption{Attention weight estimator employing different approaches to process multi-channel features. Grey vertically stacked boxes share weights.}
    \vspace{-16pt}
    \label{fig: architecture}
\end{figure}
To obtain the attention weights, a self-attention-based \gls{dnn} (a transformer encoder~\cite{vaswaniAttentionAllYou2017}) is employed, i.e.,
\begin{align}
    \left\{ \mathbf{a}_t \right\}_{t=1}^T &= \mathrm{DNN} \left( \left\{ \mathbf{i}_t \right\}_{t=1}^{T}; \ \boldsymbol{\Lambda}\right), 
\end{align}
where $\mathbf{a}_t = [(\mathbf{a}^{x}_t)\tran, \ (\mathbf{a}^{n}_t)\tran]\tran$ denotes the attention weights at time frame $t$, $\mathbf{i}_t = [(\mathbf{i}^{x}_t)\tran, \ (\mathbf{i}^{n}_t)\tran]\tran$ denotes the input features at time frame $t$, and $\boldsymbol{\Lambda}$ denotes the parameters of the \gls{dnn}\footnote{In \cite{ochiaiMaskBasedNeuralBeamforming2023}, $\mathrm{DNN}^x(\cdot)$ and $\mathrm{DNN}^n(\cdot)$ were used separately for the speech and noise components. Our preliminary experiments showed a similar or better performance at a lower computational complexity when using a single \gls{dnn}.}.
As illustrated in Fig.~\ref{fig: architecture} (top), the input features are first transformed into a time-varying embedding vector via a linear layer.
This embedding vector then passes through several \gls{mha} encoder blocks, each comprising multi-head self-attention layers and position-wise feedforward layers, which are all interconnected through residual connections.
Finally, the attention weights $\mathbf{a}_t$ are extracted from two separate speech and noise \gls{sha} layers.
In \cite{ochiaiMaskBasedNeuralBeamforming2023}, the \glspl{iscm} defined in \eqref{eq: iscm} are used as the speech and noise input features, i.e.,
\begin{align}
\label{eq: features ISCM}
    \mathbf{i}^{\nu, \mathrm{ISCM}}_{f,t} = \left[ \Re \left( \mathrm{vec}(\widehat{\boldsymbol{\Psi}}^{\nu}_{f,t})\tran \right), \ \Im \left( \mathrm{vec}(\widehat{\boldsymbol{\Psi}}^{\nu}_{f,t})\tran \right) \right]\tran \in \mathbb{R}^{2C^2},
\end{align}
where $\mathrm{vec} \left( \cdot \right)$ denotes a reshaping of a $C \times C$-dimensional matrix into a vector of length $C^2$.
The speech and noise features are concatenated along the frequency dimension, resulting in $\mathbf{i}^{\mathrm{ISCM}}_{t} \in \mathbb{R}^{4FC^2}$.

As described in Section~\ref{sec:scm}, the attention weights $\mathbf{a}_t$ control the temporal aggregation of the speech and noise \glspl{iscm}.
Although the formulation in \eqref{eq: iscm} allows for a potential application across different microphone arrays, it should be noted that the approach proposed in \cite{ochiaiMaskBasedNeuralBeamforming2023} does depend on the channel configuration.
More specifically, the training procedure used a fixed channel configuration, not accounting for channel configuration variability, while the \gls{dnn} architecture used a fixed input layer size, and the \gls{iscm} features in \eqref{eq: features ISCM} simultaneously incorporate spatial and spectro-temporal information, making them sensitive to the channel configuration considered during training.

\section{Proposed Approaches to Improve Robustness Against Channel Configuration Variations}
In this section, we propose three approaches to improve the robustness of the mask-based beamformer with \gls{asa} against channel configuration variations.

\subsection{Training With Random Channel Configurations}
\label{sec: Randomization of Microphone Selection}
To prevent the \gls{dnn} from overfitting to specific channel permutations, channel numbers, and microphone array geometries, a straightforward approach is to integrate random channel configurations into the training procedure.
Assuming that a single microphone array with $C_{\mathrm{max}}$ channels is available for training, for each minibatch a channel number $C^\prime$ is drawn from the uniform random distribution $\mathcal{U}(2, \ C_{\mathrm{max}})$.
From the available $C_{\mathrm{max}}$ channels, $C^\prime$ channels are then selected in random permutation, resulting in random microphone subarrays.

\subsection{TAC Method to Process Multi-Channel Features}
\label{sec: tac}
To accommodate a variable number of input channels in the training of the \gls{dnn} with fixed input layer size, zero-padding up to $C_{\mathrm{max}}$ channels can be applied.
However, this approach may sacrifice upper bound speech enhancement performance for robustness, since the \gls{dnn} needs to learn to deal with zero-padded input features, while also being limited to $C^\prime \leq C_{\mathrm{max}}$ channels.
To deal with this issue, we propose to employ the \gls{tac} method~\cite{luoEndtoendMicrophonePermutation2020} to process multi-channel features in the attention weight estimator, as depicted in Fig.~\ref{fig: architecture} (bottom). 
A \gls{tac} block takes as input a set of feature streams $\{\mathbf{z}_{t,c} \in \mathbb{R}^D\}^{C^\prime}_{c=1}$ with variable $C^\prime$ and a channel-independent feature dimension $D$, shares information across the streams in a non-linearly transformed space, and outputs a set of modified feature streams $\{\tilde{\mathbf{z}}_{t,c} \in \mathbb{R}^D\}^{C^\prime}_{c=1}$.
We adopt the efficient TAC implementation from \cite{yoshiokaVarArrayArrayGeometryAgnosticContinuous2022}, obtaining the modified feature stream at time frame $t$ and channel $c$ as:
\begin{equation}
\tilde{\mathbf{z}}_{t,c} = \left[\mathrm{ReLU}(\mathbf{L}_1 \mathbf{z}_{t,c})\tran, \ \frac{1}{C^\prime} \sum_{\mu=1}^{C^\prime} \mathrm{ReLU}(\mathbf{L}_2 \mathbf{z}_{t,\mu})\tran\right]^\mathsf{T},
\label{eq:tac}
\end{equation}
where $\mathbf{L}_1 \in \mathbb{R}^{(\nicefrac{D}{2}) \times D}$ and $\mathbf{L}_2 \in \mathbb{R}^{(\nicefrac{D}{2}) \times D}$ denote trainable linear transforms shared across all channels.
The modified feature streams contain channel-specific information as well as information affected by all channels in a permutation-invariant fashion due to the combination of weight sharing and the application of the permutation-invariant averaging operation.

The \gls{tac} method is integrated into the attention weight estimator by interleaving \gls{tac} blocks with $C^\prime$ parallel \gls{mha} encoder blocks sharing the same parameters (see Fig.~\ref{fig: architecture}, bottom).
After $N$ stacks of parallel interleaved \gls{tac} blocks and \gls{mha} encoder blocks, the streams are averaged and passed to the final \gls{sha} speech and noise encoder blocks.
This integration enables handling a varying channel number $C^\prime$ (even $C^\prime > C_{\mathrm{max}}$) and ensures invariance to the channel permutation.
This significantly enhances the flexibility and applicability of the attention weight estimator across diverse channel configurations, without necessitating modifications in the \gls{dnn} architecture or hyperparameters.

{
\setlength{\tabcolsep}{5.25pt} 
\begin{table*}[t]
\centering
\footnotesize
\caption{Mean \acrshort{pesq} and \acrshort{sdr} values for the noisy mixtures, a mask-based \gls{mvdr} beamformer with recursive smoothing using a fixed forgetting factor, and the mask-based \gls{mvdr} beamformer with \gls{asa} employing different attention weight estimators, evaluated on datasets corresponding to a matched condition and various mismatched conditions.
}
\begin{tabular}{lllllrrrrrrrrrrrr}
\toprule
\multicolumn{5}{l}{} & \multicolumn{2}{c}{matched} & \multicolumn{8}{c}{mismatched in terms of} \\
\cmidrule(lr){6-7} \cmidrule(lr){8-15}
\multicolumn{5}{l}{} & \multicolumn{2}{c}{} & \multicolumn{2}{c}{Permutation} & \multicolumn{2}{c}{Number} & \multicolumn{2}{c}{Geometry} & \multicolumn{2}{c}{Number \& Geom.} \\
\cmidrule(lr){8-9} \cmidrule(lr){10-11} \cmidrule(lr){12-13} \cmidrule(lr){14-15}
 & & config. & features & use TAC & PESQ & SDR & PESQ & SDR & PESQ & SDR & PESQ & SDR & PESQ & SDR\\
\midrule
 1 & \multicolumn{1}{l}{mixture} & --- & --- & --- &1.37 & 5.19 & 1.37 & 5.19 & 1.37 & 5.19 & 1.38 & 3.12 & 1.38 & 3.12\\
 2 & \multicolumn{1}{l}{recursive \gls{mvdr}} & --- & --- & --- & 2.04 & 10.18 & 2.04 & 10.18 & 1.73 & 9.05 & 2.00 & 8.94 & 1.73 & 7.40\\
 3 & baseline~\cite{ochiaiMaskBasedNeuralBeamforming2023} & fixed & ISCM & False & \textbf{2.64} & 16.34 & 2.31 & 13.72 & 1.84 & 10.66 & 2.19 & 11.32 & 1.71 & 7.39\\
\cmidrule(lr){1-15}
 4 & \multirow{2}{*}{proposed} & fixed & mag-IPD & False & 2.57 & 16.27 & 2.40 & 14.70 & 1.84 & 10.71 & 2.15 & 11.01 & 1.77 & 8.34 \\
 5 &  & fixed & mag-IPD & True & 2.62 & \textbf{16.39} & \textbf{2.62} & \textbf{16.39} & 2.05 & 12.55 & 2.20 & 11.84 & 1.87 & 9.25\\
\cmidrule(lr){1-15}
 6 & \multirow{3}{*}{proposed} & random & ISCM & False & 2.42 & 14.37 & 2.42 & 14.36 & 1.96 & 11.86 & 2.18 & 11.55 & 1.93 & 10.02\\
 7 &  & random & mag-IPD & False & 2.53 & 15.85 & 2.52 & 15.86 & 2.32 & 14.07 & 2.18 & 11.82 & 1.99 & 10.54\\
 8 &  & random & mag-IPD & True & 2.59 & 16.02 & 2.59 & 16.02 & \textbf{2.34} & \textbf{14.15} & \textbf{2.21} & \textbf{12.35} & \textbf{2.03} & \textbf{11.34}\\
\bottomrule
\end{tabular}
\vspace{-11pt}
\label{tab: detailed results combined}
\end{table*}}

\subsection{Input Features}
Due to the definition of the \gls{iscm} in \eqref{eq: iscm}, the input features in \eqref{eq: features ISCM} simultaneously encode \glspl{ild} as well as \glspl{ipd} and hence strongly depend on the microphone array geometry.
In addition, the $4FC^2$-dimensional features $\mathbf{i}^{\mathrm{ISCM}}_{t}$ below \eqref{eq: features ISCM} are incompatible with the \gls{tac} method, since it requires channel-wise feature streams with a channel number-independent feature dimension.

To address these issues, we propose to adopt alternative channel-wise feature streams
(denoted as mag-\gls{ipd} features), defined as:
\begin{equation}
    \label{eq: features mag-IPD}
    \mathbf{i}^{\nu, \text{mag-IPD}}_{f,t,c} = \left[|\widehat{\nu}_{f,t,c}|^2, \ \cos\left(\widehat{\delta}_{f,t,c}\right), \ \sin\left(\widehat{\delta}_{f,t,c}\right)\right]\tran,
\end{equation}
where $\widehat{\delta}_{f,t,c} = \angle \widehat{\nu}_{f,t,c} - \angle \widehat{\sigma}_{f,t}$ denotes the difference between the unwrapped phases of the masked \gls{stft} coefficients $\widehat{\nu}_{f,t,c} = m^\nu_{f,t} y_{f,t,c}$ and the channel-averaged masked \gls{stft} coefficients $\widehat{\sigma}_{f,t} = \frac{1}{C} \sum^{C}_{c=1} \widehat{\nu}_{f,t,c}$, and $\cos$ and $\sin$ have been applied to result in a smooth phase representation.
The features proposed in \eqref{eq: features mag-IPD} differ from those in \cite{yoshiokaVarArrayArrayGeometryAgnosticContinuous2022}, which used $|\widehat{\sigma}_{f,t}|^2$ instead of $|\widehat{\nu}_{f,t,c}|^2$ and the phase component instead of its cosine and sine, yielding a worse performance in our preliminary experiments.
Concatenating the speech and noise features along the frequency dimension yields $C$ streams of $6F$-dimensional features $\mathbf{i}^{\text{mag-IPD,TAC}}_{t,c}$ (see Fig.~\ref{fig: architecture}, bottom).
We hypothesize that these features are less sensitive to the channel configuration than the features in \eqref{eq: features ISCM} because they do not explicitly depend on channel pairs and they effectively separate the channel configuration-dependent \gls{ipd} information from magnitude information, which is less influenced by the channel configuration.
In addition, we employ the proposed features with the conventional attention weight estimator, in which case we concatenate the speech and noise features along the frequency and channel dimensions, yielding $6FC$-dimensional features $\mathbf{i}^{\text{mag-IPD,cat}}_{t}$ (see Fig.~\ref{fig: architecture}, top).

\section{Experiments}
\subsection{Datasets}
\label{sec: Datasets}
\begin{figure}[t]
    \centering
    \hspace{30pt}
    \begin{subfigure}{61.93pt}
\begingroup%
  \makeatletter%
  \providecommand\color[2][]{%
    \errmessage{(Inkscape) Color is used for the text in Inkscape, but the package 'color.sty' is not loaded}%
    \renewcommand\color[2][]{}%
  }%
  \providecommand\transparent[1]{%
    \errmessage{(Inkscape) Transparency is used (non-zero) for the text in Inkscape, but the package 'transparent.sty' is not loaded}%
    \renewcommand\transparent[1]{}%
  }%
  \providecommand\rotatebox[2]{#2}%
  \newcommand*\fsize{\dimexpr\f@size pt\relax}%
  \newcommand*\lineheight[1]{\fontsize{\fsize}{#1\fsize}\selectfont}%
  \ifx\svgwidth\undefined%
    \setlength{\unitlength}{61.92881715bp}%
    \ifx\svgscale\undefined%
      \relax%
    \else%
      \setlength{\unitlength}{\unitlength * \real{\svgscale}}%
    \fi%
  \else%
    \setlength{\unitlength}{\svgwidth}%
  \fi%
  \global\let\svgwidth\undefined%
  \global\let\svgscale\undefined%
  \makeatother%
  \begin{picture}(1,0.90716959)%
    \lineheight{1}%
    \setlength\tabcolsep{0pt}%
    \put(0,0){\includegraphics[width=\unitlength,page=1]{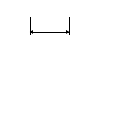}}%
    \put(0.38284057,0.80303128){\color[rgb]{0,0,0}\makebox(0,0)[t]{\lineheight{1.25}\smash{\begin{tabular}[t]{c}$d_1$ = 10 cm\end{tabular}}}}%
    \put(0,0){\includegraphics[width=\unitlength,page=2]{chime.pdf}}%
    \put(0.56909906,0.35378498){\color[rgb]{0,0,0}\makebox(0,0)[t]{\lineheight{1.25}\smash{\begin{tabular}[t]{c}$d_2$ = 19 cm\end{tabular}}}}%
    \put(0,0){\includegraphics[width=\unitlength,page=3]{chime.pdf}}%
  \end{picture}%
\endgroup%

        \caption{CHiME-3~\cite{barkerThirdCHiMESpeech2015}}
        \label{fig: geometry: chime}
    \end{subfigure}
    \hfill
    \begin{subfigure}{63.75pt}
\begingroup%
  \makeatletter%
  \providecommand\color[2][]{%
    \errmessage{(Inkscape) Color is used for the text in Inkscape, but the package 'color.sty' is not loaded}%
    \renewcommand\color[2][]{}%
  }%
  \providecommand\transparent[1]{%
    \errmessage{(Inkscape) Transparency is used (non-zero) for the text in Inkscape, but the package 'transparent.sty' is not loaded}%
    \renewcommand\transparent[1]{}%
  }%
  \providecommand\rotatebox[2]{#2}%
  \newcommand*\fsize{\dimexpr\f@size pt\relax}%
  \newcommand*\lineheight[1]{\fontsize{\fsize}{#1\fsize}\selectfont}%
  \ifx\svgwidth\undefined%
    \setlength{\unitlength}{63.74550236bp}%
    \ifx\svgscale\undefined%
      \relax%
    \else%
      \setlength{\unitlength}{\unitlength * \real{\svgscale}}%
    \fi%
  \else%
    \setlength{\unitlength}{\svgwidth}%
  \fi%
  \global\let\svgwidth\undefined%
  \global\let\svgscale\undefined%
  \makeatother%
  \begin{picture}(1,0.88131614)%
    \lineheight{1}%
    \setlength\tabcolsep{0pt}%
    \put(0,0){\includegraphics[width=\unitlength,page=1]{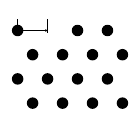}}%
    \put(0.24490467,0.7595077){\color[rgb]{0,0,0}\makebox(0,0)[t]{\lineheight{1.25}\smash{\begin{tabular}[t]{c}$d$ = 5 cm\end{tabular}}}}%
    \put(0,0){\includegraphics[width=\unitlength,page=2]{demand.pdf}}%
    \put(0.12202209,0.48500423){\color[rgb]{0,0,0}\makebox(0,0)[t]{\lineheight{1.25}\smash{\begin{tabular}[t]{c}$d$\end{tabular}}}}%
    \put(0,0){\includegraphics[width=\unitlength,page=3]{demand.pdf}}%
  \end{picture}%
\endgroup%

        \caption{DEMAND~\cite{thiemannDiverseEnvironmentsMultichannel2013}}
        \label{fig: geometry: demand}
    \end{subfigure}
    \hspace{30pt}
    \caption{Considered microphone array geometries. Grey circles denote the reference and white circles denote unused microphones.}
    \vspace{-16pt}
    \label{fig: geometry}
\end{figure}
To evaluate the effectiveness of the proposed approaches, we constructed datasets of simulated moving speakers in noisy conditions using speech signals from the \gls{wsj0} corpus~\cite{paulDesignWallStreet1992} and noise recordings from the CHiME-3~\cite{barkerThirdCHiMESpeech2015} and DEMAND~\cite{thiemannDiverseEnvironmentsMultichannel2013} corpora.
We constructed two datasets with different microphone array geometries (illustrated in Fig.~\ref{fig: geometry}), both with a sampling frequency of \SI{16}{\kilo\hertz}.
Similarly as in \cite{ochiaiMaskBasedNeuralBeamforming2023}, we simulated speakers moving on a linear trajectory with constant speed using the gpuRIR tool~\cite{diaz-guerraGpuRIRPythonLibrary2021} by generating \glspl{rir} at 128 positions on a line, with room width and depth uniformly drawn from the set of \ \{\SI{3.0}{\meter}, \SI{3.5}{\meter}, \SI{4.0}{\meter}, \SI{4.5}{\meter}, \SI{5.0}{\meter}\}, room height equal to \SI{2.5}{\meter}, reverberation time $T_{60}$ drawn uniformly between \SIlist{0.1; 0.3}{\second}, and the microphone array randomly placed in the room.
We added the speech signal convolved with the simulated \glspl{rir} and the recorded noise signals at \glspl{snr} between \SIlist{2; 8}{\decibel}.

The first dataset consists of simulated utterances based on the \gls{wsj0} speech and CHiME-3 noise signals, resulting in a maximum number of $C_{\mathrm{max}} = 5$ channels available for training (excluding the rear-facing second channel).
This dataset was used for training, development, and evaluation.
The second dataset consists of simulated utterances based on the \gls{wsj0} speech and DEMAND noise signals, resulting in 16 available channels.
This dataset was only used for evaluation.

During evaluation, we considered a matched condition and several mismatched conditions.
``matched'' represents the CHiME-3-based evaluation dataset with a fixed channel permutation and the channel number $C^\prime = C_{\mathrm{max}} = 5$, similar to the fixed training condition.
To evaluate a mismatch in terms of the channel permutation, we randomly permuted the channels from the CHiME-3-based evaluation dataset.
To evaluate a mismatch in terms of the channel number, we selected the first $C^\prime=3$ channels from the CHiME-3-based evaluation dataset.
To evaluate a mismatch in terms of the microphone array geometry, we randomly selected $C^\prime=5$ channels from the DEMAND-based evaluation dataset.
This procedure allows for diverse microphone array geometries, e.g., including linear, triangular, rectangular, and trapezoidal shapes, some of which are not realizable with the CHiME-3 microphone array used for training (see Fig.~\ref{fig: geometry}).
To evaluate a mismatch in terms of both the channel number and the microphone array geometry, we randomly selected $C^\prime=3$ channels from the DEMAND-based evaluation dataset.
In all evaluation conditions, the reference channel was chosen as depicted in Fig.~\ref{fig: geometry}.
We created \num{30000}, \num{2000}, and \num{2000} noisy utterances for training, development, and each evaluation dataset, respectively.

\subsection{Settings}
\label{sec: simulations: settings}
We mostly followed the experimental settings presented in \cite{ochiaiMaskBasedNeuralBeamforming2023} to increase comparability with the associated results.
We trained the attention weight estimator in an end-to-end manner, utilizing the scale-dependent \gls{snr} loss function~\cite{lerouxSDRHalfbakedWell2019} at the output of the mask-based beamformer (see Fig.~\ref{fig: overview}), with the reverberant clean speech component at the reference microphone as the target signal.
During training, we used oracle Wiener-like time-frequency masks \cite{erdoganPhasesensitiveRecognitionboostedSpeech2015} to compute the \glspl{iscm} in \eqref{eq: iscm} and optimized only the trainable parameters of the attention weight estimator.
During the evaluation, we used a time-frequency mask estimator based on a temporal convolutional network architecture \cite{luoConvTasNetSurpassingIdeal2019}. 
For the attention weight and time-frequency mask estimators, we adopted the \gls{dnn} and training hyperparameters in \cite{ochiaiMaskBasedNeuralBeamforming2023}, except for using a single \gls{dnn} for both the speech and noise components (see Section \ref{sec: attention weight estimation}).
For the \gls{tac} blocks, we adopted the implementation proposed in \cite{yoshiokaVarArrayArrayGeometryAgnosticContinuous2022}, consisting of linear layers and ReLU activations (cf. \eqref{eq:tac}). 

In addition to the mask-based \gls{mvdr} beamformer with the original \gls{asa} in \cite{ochiaiMaskBasedNeuralBeamforming2023}, we considered a mask-based \gls{mvdr} beamformer with recursive smoothing using a fixed (frequency-independent) forgetting factor that corresponds to a time constant of \SI{1.6}{\second} (tuned according to the highest \gls{sdr} values under the matched evaluation condition) as a baseline algorithm~\cite{ochiaiMaskBasedNeuralBeamforming2023,higuchiRobustMVDRBeamforming2016}.
For the \gls{stft}, we used a Hann window with a frame length of \SI{64}{\milli\second} and \SI{16}{\milli\second} shift.

We evaluated the speech enhancement performance in terms of \gls{pesq}~\cite{rixPerceptualEvaluationSpeech2001} and \gls{sdr}~\cite{vincentPerformanceMeasurementBlind2006} (allowing for distortions caused by time-invariant filters), with the reverberant clean speech component at the reference microphone as the reference signal.

\subsection{Results}
\label{sec: simulations: results}
Table \ref{tab: detailed results combined} shows the mean \gls{pesq} and \gls{sdr} values for the noisy mixtures, the mask-based beamformer with recursive smoothing (described in the previous section), and for the mask-based beamformer with \gls{asa} employing different attention weight estimators (baseline estimator in \cite{ochiaiMaskBasedNeuralBeamforming2023} and proposed estimators).
In this table, ``config'' indicates whether the channel permutation and number were fixed or randomized during training (see Section \ref{sec: Randomization of Microphone Selection}); ``features'' represents the utilized input features, either the \gls{iscm} features in \eqref{eq: features ISCM} or the proposed mag-\gls{ipd} features in \eqref{eq: features mag-IPD}; ``use TAC'' indicates whether \gls{tac} was employed or not.
We evaluated these beamformers both under a matched condition as well as under various mismatched conditions described in Section \ref{sec: Datasets}.

The results in Table \ref{tab: detailed results combined} show that under all conditions both the mask-based beamformer with recursive smoothing as well as the mask-based beamformer with \gls{asa} (for all attention weight estimators) substantially improve the \gls{pesq} and \gls{sdr} values compared to the noisy mixtures.
Under the matched condition, it can be observed that models trained with a fixed channel configuration (rows 3-5) achieve the highest \gls{pesq} and \gls{sdr} values.
This is expected as these models can exploit the specific spatial information seen during training, representing an upper bound in performance.

Under mismatched conditions, the baseline model (row 3) shows notable performance degradation, particularly in terms of channel number and microphone array geometry.
The model employing mag-\gls{ipd} features (row 4) exhibits a similar performance as the baseline model in most conditions, except for a reduced performance drop under the channel permutation mismatch.
The model employing \gls{iscm} features with randomized training configurations (row 6) demonstrates similar robustness across mismatched conditions as the model in row 4, albeit with a worse performance under the matched condition, highlighting a trade-off between robustness and upper bound performance.
The incorporation of mag-\gls{ipd} features and the \gls{tac} method (row 5) further mitigates performance drops across all mismatch conditions, completely alleviating the drop under the channel permutation mismatch while maintaining strong matched condition performance.
The model combining mag-\gls{ipd} features, \gls{tac}, and randomized training configurations (row 8) achieves the most consistent high performance, performing similarly as the best model under the matched condition and the channel permutation mismatch conditions (row 5), as well as outperforming all models under channel number and microphone array geometry mismatches.
The results clearly show that the combination of training with random channel configurations, employing the \gls{tac} method, and using the mag-\gls{ipd}-based input features resulted in a significantly higher speech enhancement performance compared to the baseline model~\cite{ochiaiMaskBasedNeuralBeamforming2023} (significance determined using a two-sided T-test with Bonferroni correction).

It should be emphasized that the evaluation included diverse microphone array geometries by randomly selecting channels from the DEMAND-based evaluation dataset, i.e., ``Geometry'' and ``Number \& Geom.'' in Table \ref{tab: detailed results combined}.
Hence, the results show that the mask-based beamformer with \gls{asa} using the combination of all proposed approaches can perform noise reduction for moving speakers and arbitrary microphone arrays, consistently outperforming the mask-based beamformer with recursive smoothing and the baseline mask-based beamformer with the original \gls{asa}.

\section{Conclusion}
In this paper, we proposed several approaches to improve the robustness of the mask-based beamformer with \gls{asa} against channel configuration variations.
These approaches include the integration of random channel configurations during training, employing the \gls{tac} method to process multi-channel features (allowing for any channel number and enabling permutation invariance), as well as using mag-\gls{ipd} features that are robust against channel configuration variations.
Experiments using the CHiME-3 and DEMAND datasets suggest that the mask-based beamformer with \gls{asa} integrating the proposed approaches can perform noise reduction for moving speakers and arbitrary microphone arrays.
Future research will extend this investigation to explore more diverse channel configurations during training and evaluation as well as address the computational complexity of the proposed \gls{tac} integration.

\clearpage

\bibliographystyle{IEEEtran}
\bibliography{refs_bibtex}

\end{document}